\documentclass[12pt]{article}
\usepackage[dvips]{color}
\usepackage{amsmath}
\usepackage{graphicx}
\usepackage{subfigure}
\usepackage{epsfig}
\usepackage{amsmath}
\usepackage{cite}
\usepackage{color}
\usepackage{subfigure}

\begin{document}
\begin{center}
\Large{\bf Investigation of hidden symmetry in $AdS_{5}$ black hole with Heun equation}\\
\small \vspace{1cm} {\bf Jafar Sadeghi $^{\star}$\footnote {Email:~~~pouriya@ipm.ir}}, \quad  {\bf Mohammad Ali S. Afshar $^{\star}$\footnote {Email:~~~m.a.s.afshar@gmail.com}}, \quad
 {\bf Mohammad Reza Alipour $^{\star}$\footnote {Email:~~~mr.alipour@stu.umz.ac.ir}}, \\ \vspace{0.2cm} {\bf Saeed Noori Gashti$^{\star,\dag}$\footnote {Email:~~~saeed.noorigashti@stu.umz.ac.ir}}

\vspace{0.5cm}$^{\star}${Department of Physics, Faculty of Basic
Sciences,\\
University of Mazandaran
P. O. Box 47416-95447, Babolsar, Iran}\\
\vspace{0.2cm}$^{\dag}${School of Physics, Damghan University, P. O. Box 3671641167, Damghan, Iran}
\small \vspace{0.15cm}

\end{center}
\begin{abstract}
Investigating the existence of algebra and finding hidden symmetries in physical systems is one of the most important aspects for understanding their behavior and predicting their future. Expanding this unique method of study to cosmic structures and combining past knowledge with new data can be very interesting and lead to discovering new ways to analyze these systems. However, studying black hole symmetries always presents many complications and sometimes requires computational approximations.
For example, checking the existence of Killing vectors and then calculating them is not always an easy task. It becomes much more difficult as the structure and geometry of the system become more complex. In this work, we will show that if the wave equations with a black hole background can be converted in the form of general Heun equation, based on its structure and coefficients, the algebra of the system can be easily studied, and computational and geometrical complications can be omitted. For this purpose, we selected two $AdS_5$ black holes: Reissner-Nordström (R-N) and Kerr, and analyzed the Klein-Gordon equation with the background of these black holes.
Based on this concept, we observed that the radial part of the R-N black hole and both the radial and angular parts of the Kerr black hole could be transformed into the general form of the Heun equation. As a result, according to the algebraic structure that governs the Heun equation and its coefficients, one can easily achieve generalized $sl(2)$ algebra.\\

\end{abstract}
\newpage
\tableofcontents

\section{Introduction}
It must be admitted that throughout the history of physics, conservation laws are the most basic tool of a physicist to study a dynamic and playful universe. Undoubtedly, the first step in studying such a system is to check for the existence of constants and stability within it. In this way, the existence of "symmetries" and "periodicity" is the most important data that these systems provide, hidden from the effects of time and without any changes.
When symmetry and periodicity are raised, the forced relationship of system elements together (algebra) and the geometric rules of this repetition (topology) play a more fundamental role.\\ Sometimes this role even becomes so powerful that it dominates the basic cognitive tools of a physicist and, as a result, makes decisions about the physics of the system based only on geometry, topology, and algebra, regardless of conventional methods. The meaning of these words is more tangible for particle physicists and cosmologists because we are not dealing with a crystal where everything can be tested and verified.\\
This is a fact! All we have is a black hole in infinity, and we are so far away that all our seemingly advanced measuring tools seem practically unusable or useless. This is where the value of long-range measurement tools such as algebra, which can reproduce system generators based on inductive reasoning, geometry, and comparison of symmetric behaviors, is determined. Then, with the help of the golden key of Noter's theorem, the relevant constants and currents will be easily available to us so that we can safely make decisions about the behaviors and results of the system.\\ But there are two ways to use the algebraic method: first of all, like a mathematician, regardless of whether the results are meaningful or meaningless from a physical point of view, consider the black hole as a mathematical entity and be satisfied only with obtaining logical results. Or be ourselves, a physicist that prioritizes conceptualization and subjectivity and use algebra only as a study tool.
We have chosen the second approach in this article.\\
Also the study of hidden symmetries in black holes is not a new concept and is often based on analyzing the wave field equation with a black hole background and then identifying the Killing vectors and the Killing-Yano tensor. This method has been used in the basic work of Castro, Mallory, and Strominger for the Kerr-Newman black hole \cite{1}, as well as in subsequent works such as \cite{2,3,4,5}, which were based on Castro and Maloney's approach for R-N black holes and 5D black holes.
It is well-known that constructing conformal coordinates and generating the Killing vectors can be a challenging task when dealing with complex black hole geometries.\\ {We propose that if it is possible to convert the obtained differential equations into general form of Heun’s equation, perhaps using this method to identify the algebra of hidden symmetries in field-particle equations with a black hole background is more appropriate, since it can provide a direct and faster way to reveal the algebra and hidden symmetries of black holes, which we will show that for the several different examples.\\
Prior to concluding this section, it is imperative to underscore that the transformation of various physical systems' differential equations into recognized forms of Heun’s equations has elevated the significance of these equations in propelling the exploration of physical systems \cite{13,14,15,16,17,18,19,21,23,24}. Various topics, such as the Schrödinger equation with anharmonic potential, the Teukolsky linear perturbation theory for the Schwarzschild and Kerr metrics, traversable wormholes, quantum Rabi models, the perturbations of rotating and non-rotating black holes, the gravitational lensing by black holes and wormholes, the quantum tunneling, the Hawking radiation, the quasinormal modes, Klein-Gordon, Dirac, Schrödinger, Maxwell, and Einstein equations in various space-times, etc., have motivated many researchers in recent decades to solve, interpret and justify differential equations arising from physical systems that turn into different forms of Heun’s equations.\cite{30,31,32,33,34,3401,3402,3403,35,36,37,38,3801,38011,3802,3803}.\\
Now, based on all the explanations above, we will organize the article in the following manner.In sections 2 and 3, we provide an explanation of algebra, Heun's equation, and work motivation. In section 4, we will give a general overview of Heun equations. In section 5, we investigate the scalar field equation in the background of the $RN-AdS_{5}$ black hole and the Heun equation of its radial part by introducing the Klein–Gordon particle around it. We identify its operators and the properties of displacement and consider the generalized $sl(2)$ algebra for the $RN-AdS_{5}$ black hole. In section 6, we will follow the same process for the $Kerr-AdS_{5}$ black hole. Finally, we describe the results in section 7.
\section{Warming the mind}
Maloney, Castro, and Strominger \cite{1} showed at the beginning of the last decade that the physical structure of a Kerr-Newman black hole can be strongly influenced by the algebra of conformal field theory. They had their reasons for this concept: for example, the wave equation around a black hole in a certain limit ("near region") can be calculated as a quadratic Casimir of an $sl(2, R)\times sl(2, R)$ algebra. They were able to calculate the entropy of the black hole using the concept of central charge near the extreme limit points and showed that the absorption cross section for scalars in the near-region can be reinterpreted as a standard finite temperature CFT absorption cross section \cite{3}.
This new path, and the checking of its validity, prompted others to move to other black holes as R-N to study the hidden symmetries of this black hole \cite{2}. But the main point in all of these works was that they all used the conventional method during this path.
It means Maloney's method! The basis of Castro and Maloney's work was that they considered the Klein-Gordon wave equation with the background of the Kerr black hole. Then, they obtained the radial and angular equations by separating the variables. Although they pointed out that these equations are solvable according to the Heun form, they went on to construct conformal coordinates and consider generators for the vector field on these coordinates. Then, they showed that these generators follow the $sl(2,R)$ algebra.
In other words, they rely on the conventional method of using Killing vectors to identify symmetry. However, in this work, we demonstrate that it is not necessary to construct a new vector field to obtain the hidden algebras of a black hole. Instead, it can be achieved as soon as the wave equation appears in the form of Heun. Depending on the type of algebraic structure of the equation and its coefficients, we can directly proceed to generate and reconstruct the algebra, and then verify its validity by checking the Lie bracket.
\section{Why Heun?}
Bender and Dunne \cite{6} showed that there is a logical correspondence between orthogonal polynomials and quasi-exactly solvable (QES) models. In other words, they demonstrated that the wave function in quantum mechanics can act as a generator for orthogonal polynomials expressed in terms of energy. Typically, the algebraic approach is used to analyze QES models, where the Hamiltonian is represented as a non-linear combination of Lie algebra generators. Based on this concept, Turbiner claimed in \cite{7} that most QES models are influenced by the $sl(2,Q)$ algebra (although it was later discovered that some QES models do not necessarily follow the $sl(2)$ algebra).
He was able to show that for different models of specific potentials of this class of Schrödinger equations, these equations can be written using the structure of polynomials and Hamiltonian reconstruction in terms of generators of the $sl(2)$ algebra. Therefore, it can be said that due to the basic and structural similarity of the Schrödinger and Klein-Gordon equations, differential equations resulting from physical systems whose solutions are somehow the result of orthogonal polynomials may lead us to the existence of a hidden algebra in the physical system. Heun's equation is one such equation. For this reason, in this article, this feature can be used as a practical, easy, and short way to get rid of the construction of Killing vectors, which usually result from long paths with complex computational problems.
The important point is that Castro and Maloney's method confirms the result obtained based on this method.

\section{Overview of Heun Equation}
{In 1889, Karl M. W. L. Heun introduced a new equation \cite{8}, a linear ordinary differential equation of second-order type ,which has four regular singular points at zero, one, an arbitrary point between zero and one, and infinity on the complex plane.In fact, this equation is a generalization of the hypergeometric equation, which undoubtedly has found many applications in various branches of science, including in physics, especially in general relativity and astrophysics. To study about this equation and its properties, we can refer to the book of Ronveaux \cite{9}, which is considered to be comprehensive in this field. \\
Based on this, the general form of Heun's equation will be as follows
\begin{equation}\label{1}
\begin{split}
y''(x)+\bigg[\frac{\eta}{x}+\frac{\delta}{x-1}+\Sigma^{k}_{i=1}\frac{\epsilon_{i}}{x-a_{i}}\bigg]y'(x)+\frac{\mu\nu x^{k}+\Sigma^{k}_{i=1}p_{i}x^{i-1}}{x(x-1)\Pi^{k}_{i=1}(x-a_{i})}y(x)=0,
\end{split}
\end{equation}
where each singularity $a_{i}$ has the characteristic $(0, 1-\epsilon_{i}), (0, 1-\eta), (0, 1-\delta)$ as well as $(\mu, \nu)$ at $x=0, x=1$ and $x=\infty$. Therefore, considering the Fuchsian relation, we will have,
\begin{equation}\label{2}
\begin{split}
\alpha+\beta+1=\eta+\delta+\Sigma^{k}_{i=1}\epsilon_{i}.
\end{split}
\end{equation}
So, the Heun equation \eqref{1} in the case of $k = 1$ is as follows,
\begin{equation}\label{3}
\begin{split}
y''(x)+\bigg[\frac{\eta}{x}+\frac{\delta}{x-1}+\frac{\epsilon_{1}}{x-a}\bigg]y'(x)+\frac{\alpha\beta x+p_{1}}{x(x-1)(x-a)}y(x)=0,
\end{split}
\end{equation}
furthermore, we can rewrite Heun's differential equation \eqref{3} in a general form by using the functions $g_1(x)$, $g_2(x)$, and $g_3(x)$,
\begin{equation}\label{4}
g_1(x)y''(x)+g_2(x)y'(x)+g_3(x)y(x)=0,
\end{equation}
where
\begin{equation}\label{5}
\begin{split}
&g_1(x)=a_0 x^3+a_1x^2+a_2x+a_3 ,\\
&g_2(x)=a_4x^2+a_5x+a_6,\\
&g_3(x)=a_7x+a_8 .
\end{split}
\end{equation}
With $a_{i (i=0,1,...,8)}\in {R}$, we can express the coefficients of $a_i$ by comparing equations \eqref{3} and \eqref{4} in terms of the parameters of Heun's equation. Additionally, as it shown in \cite{2401,24a,24b}, we can rewrite the Heun equation using equations \eqref{5} and \eqref{4} in terms of operators $P^0$, $P^-$, $P^+$, and $F(x\frac{d}{dx})$, such that they follow the generalized $sl(2)$ algebra and the following commutation relation holds for them,
\begin{equation}\label{6}
\begin{split}
[P^0, P^{\pm}]=\pm \alpha P^{\pm} \qquad  [P^+, P^-]=FP^0 ,
\end{split}
\end{equation}
where
\begin{equation}\label{7}
\begin{split}
&P^0=\alpha x \frac{d}{dx}+\beta\\
&P^+=a_0 x^3 \frac{d^2}{dx^2}+a_4x^2\frac{d}{dx}+a_7x \\
&P^-=a_2x \frac{d^2}{dx^2}+a_6\frac{d}{dx}  \\
&F(x\frac{d}{dx})=a_1x^2\frac{d^2}{dx^2}+a_5 x\frac{d}{dx}+a_8 .\\
\end{split}
\end{equation}

Also note that we have the following conditions to establish equations \eqref{6} and \eqref{7},
\begin{equation}\label{77}
\begin{split}
&\alpha =-\frac{4 a_0 a_2}{a_1},   \hspace{2cm}   \beta =-\frac{a_6 a_7}{a_8}\\
& 2 \alpha  a_1+\alpha  a_5+a_1 \beta =-\left(6 a_0 a_2+3 a_4 a_2+3 a_0 a_6\right)\\
& \alpha  a_5+\alpha  a_8+a_5 \beta =-\left(2 a_2 a_4+2 a_6 a_4+2 a_2 a_7\right) . \\
\end{split}
\end{equation}
The calculations related to equation \eqref{77} are in Appendix A. Note that the $P^0$ form is significant because it exhibits the symmetry algebra of the mode. By considering the $P^0$ form in relation \eqref{7}, we can establish the generalized $sl(2)$ algebra. Additionally, $\alpha$ and $\beta$ are special parameters that establish the relation \eqref{6}. In the following, we will discuss the existence of scalar field symmetry in the background of $RN-AdS_{5}$ and $Kerr-AdS_{5}$ black holes.
\section{$RN-AdS_{5}$ black hole}
First, we introduce the $RN-AdS_{5}$ black hole. The line element of the $RN-AdS_{5}$ black hole is as follows\cite{25},
\begin{equation}\label{8}
\begin{split}
ds^{2}=-f(r)dt+\frac{1}{f(r)}dr^{2}+r^{2}d\Omega_{3}^{2},
\end{split}
\end{equation}
where $d\Omega_{3}^{2}$ represents the metric of the unit three-sphere. Additionally, the function $f(r)$ is defined as follows,
\begin{equation}\label{9}
\begin{split}
f(r)=1-\frac{M}{r^{2}}+\frac{Q^{2}}{r^{4}}+r^{2}=\frac{\Delta_{r}}{r^{4}}=\frac{(r^{2}-r_{+}^{2})(r^{2}-r_{-}^{2})(r^{2}-r_{0}^{2})}{r^{4}},
\end{split}
\end{equation}
where $M$ and $Q$ are the black hole's mass and charge, respectively, and the AdS radius is $L=1$. There is a correspondence between the roots $\Delta_{r}$ and the black hole's Killing horizons, whose radial positions will be parameterized by the largest real root of $r_{+}$, corresponding to the outer horizon. The roots are as follows,
\begin{equation}\label{10}
\begin{split}
&r_{-}^{2}=\frac{1}{2}\bigg[-1-r_{+}^{2}+\sqrt{(1+r_{+}^{2})^{2}+\frac{4Q^{2}}{r_{+}^{2}}}\bigg],\\
&r_{0}^{2}=\frac{1}{2}\bigg[-1-r_{+}^{2}-\sqrt{(1+r_{+}^{2})^{2}+\frac{4Q^{2}}{r_{+}^{2}}}\bigg].
\end{split}
\end{equation}
One can use the Klein-Gordon equation in curved space to derive the field equation near the black hole \cite{25},
\begin{equation}\label{11}
\begin{split}
\frac{1}{\sqrt{-g}} D_{\mu}(\sqrt{-g}g^{\mu\nu}D_{\nu})\Psi-m^2\Psi=0 ,
\end{split}
\end{equation}
where $D_{\mu}=\nabla_{\mu}-iqA_{\mu}$, $A_{\mu}=(-\sqrt{3}Q/2r^2,0,0,0,0)$, and $m$ and $q$ are the electromagnetic potential, mass, and charge of the scalar field, respectively. Also, we can relate the mass of the scalar field in equation \eqref{11} to the CFT dimension, such that $m^2=\Delta(\Delta-4)$.
In \cite{25}, the differential equation for the radial part of equation \eqref{11} for the scalar field near the $RN-AdS_5$ black hole was obtained by defining $z=\frac{r^2-r_-^2}{r^2-r_0^2}$,
\begin{equation}\label{12}
\begin{split}
F''(z)+\left[\frac{1-\theta_-}{z}+\frac{1-\theta_+}{z-z_0}+\frac{\Delta -1}{z-1}\right]F'(z)+\left[\frac{k_1k_2}{z(z-1)}-\frac{z_0(z_0-1)K_0}{z(z-z_0)(z-1)}\right]F(z)=0 ,
\end{split}
\end{equation}
where
\begin{equation}\label{13}
\begin{split}
&k_1=\frac{1}{2}(\theta_++\theta_--\theta_0-\Delta)\\
&k_2=\frac{1}{2}(\theta_++\theta_-+\theta_0-\Delta)\\
&z_0=\frac{r_+^2-r_-^2}{r_+^2-r_0^2}\\
&K_0=-\frac{[(\theta_++\theta_-+1)^2-1-\theta_0^2]}{4z_0}-\frac{[(2-\Delta)^2+2(\theta_+-1)(1-\Delta)-2]}{4(z_0-1)}\\
&-\frac{\ell(\ell+2)+\Delta(\Delta-4)r_-^2-\omega^2}{4z_0(z_0-1)(r_+^2-r_0^2)},
\end{split}
\end{equation}
where $\ell$ and $\omega$ are the angular momentum quantum number  and the quasinormal frequency, respectively. Also, parameters $\{ \theta_n, n=+,-,0 \}$ are $\theta_{n}=\frac{i}{2\pi T_{n}}\left(\omega-\sqrt{3}qQ/2r_n^2\right)$.
We rewrite equation \eqref{12} as equation \eqref{4},
\begin{equation}\label{14}
\begin{split}
&(z^3-z^2(z_0+1)+zz_0)F''(z)\\
&+[(1-\theta_- -\theta_++\Delta)z^2+(\theta_++\theta_-+z_0\theta_--z_0\Delta-2)z+z_0(1-\theta_-)]F'(z)\\
&+[k_1k_2z-k_1k_2z_0-z_0k_0(z_0-1)]F(z)=0 .
\end{split}
\end{equation}
Now, by comparing relations \eqref{14}, \eqref{5}, \eqref{4}, and \eqref{3}, we can obtain values for $a_i$, $\alpha$, and $\beta$,
\begin{equation}\label{15}
\begin{split}
&a_0= 1, \hspace{0.5cm} a_1=-(1+z_0), \hspace{0.5cm} a_2=z_0, \hspace{0.5cm} a_3=0\\
&a_4=1+\Delta-\theta_+-\theta_-, \hspace{0.5cm} a_5=-2+\theta_+-z_0\Delta+\theta_-+z_0\theta_-, \hspace{0.5cm}  a_6=z_0(1-\theta_-)\\
&a_7=k_1k_2, \hspace{0.5cm}  a_8=-z_0[k_1 k_2+(z_0-1)K_0], \hspace{0.5cm} \alpha=-k_1, \hspace{0.5cm} \beta=-k_2 .
\end{split}
\end{equation}
Using relations \eqref{15} and \eqref{7}, we obtain the operators $P^+$, $P^-$, $P^0$, and $F(z\frac{d}{dz})$,
\begin{equation}\label{16}
\begin{split}
&P^0=-k_1 z \frac{d}{dz}-k_2\\
&P^+= z^3 \frac{d^2}{dz^2}+(1+\Delta-\theta_+-\theta_-)z^2\frac{d}{dz}+k_1k_2z \\
&P^-=z_0x \frac{d^2}{dz^2}+z_0(1-\theta_-)\frac{d}{dz}  \\
&F(z\frac{d}{dz})=-(1+z_0)z^2\frac{d^2}{dz^2}+(-2+\theta_+-z_0\Delta+\theta_-+z_0\theta_-)z\frac{d}{dz}+-z_0[k_1 k_2+(z_0-1)K_0].
\end{split}
\end{equation}
According to equation \eqref{77}, we obtain the conditions for establishing algebra for equation \eqref{16} as follows,
\begin{equation}\label{1616}
\begin{split}
&\alpha =\frac{4 z_0}{z_0+1},   \hspace{2cm}   \beta =\frac{\left(1-\theta _-\right) k_1 k_2}{k_1 k_2+K_0 \left(z_0-1\right)}\\
& \alpha  \theta _+-4 \alpha -\beta -z_0 \left(\alpha  (\Delta +2)+\beta -3 (\Delta +4)+3 \theta _+\right)+\theta _- \left(\alpha +(\alpha -6) z_0\right)=0\\
& \left(\theta _-+\theta _+-2\right) (\alpha +\beta )-\alpha  K_0 z_0^2 \\
& + z_0 \left(\theta _- \left(\alpha +\beta -2 \Delta +2 \theta _+-6\right)-\Delta  (\alpha +\beta -4)+2 \theta _-^2-4 \theta _+-(\alpha -2) k_1 k_2+\alpha  K_0+4\right)=0.\\
\end{split}
\end{equation}
Since the operators mentioned above satisfy the relation \eqref{6}, the radial part of the scalar field near the $RN-AdS_5$ black hole can exhibit generalized $sl(2)$ symmetry. Furthermore, the quadratic Casimir operator $C$, which is associated with all current operators of the model, can be obtained using the formula $C=\frac{1}{2}(P^-P^+ + P^+P^-)+P^0P^0$.
\section{$Kerr-AdS_{5}$ black hole}
The metric of the $Kerr-AdS_{5}$ spacetime is given by \cite{26,27,28},
\begin{equation}\label{17}
\begin{split}
&ds^{2}=-\frac{\Delta}{\rho^{2}}\bigg(dt-\frac{a_{1}\sin^{2}\theta}{\Xi_{1}}d\phi_{1}-\frac{a_{2}\cos^{2}\theta}{\Xi_{2}}d\phi_{2}\bigg)^{2}+\frac{\rho^{2}}{\Delta}dr^{2}\\
&+\frac{\rho^{2}}{\Delta_{\theta}}d\theta^{2}+\frac{\Delta_{\theta}\sin^{2}\theta}{\rho^{2}}\bigg(a_{1}dt+\frac{r^{2}+a_{1}^{2}}{\Xi_{1}^{2}}d\phi_{1}\bigg)^{2}
+\frac{\Delta_{\theta}\cos^{2}\theta}{\rho^{2}}\bigg(a_{2}dt+\frac{r^{2}+a_{2}^{2}}{\Xi_{2}^{2}}d\phi_{2}\bigg)^{2}\\
&+\frac{1+r^{2}/L^{2}}{r^{2}\rho^{2}}\bigg(a_{1}a_{2}dt-\frac{a_{2}(r^{2}+a_{1}^{2})\sin^{2}\theta}{\Xi_{1}}d\phi_{1}-\frac{a_{1}(r^{2}+a_{2}^{2})\cos^{2}\theta}{\Xi_{2}}d\phi_{2}\bigg)^{2},
\end{split}
\end{equation}
where,
\begin{equation}\label{18}
\begin{split}
&\Delta:=\frac{1}{r^{2}}(r^{2}+a_{1}^{2})(r^{2}+a_{2}^{2})(1+r^{2}/L^{2})-2M=\frac{1}{r^{2}L^{2}}(r^{2}-r_{0}^{2})(r^{2}-r_{+}^{2})(r^{2}-r_{-}^{2}),\\
&\Delta_{\theta}:=1-\frac{a_{1}^{2}}{L^{2}}\cos^{2}\theta-\frac{a_{2}^{2}}{L^{2}}\sin^{2}\theta,\\
&\rho^{2}:=r^{2}+a_{1}^{2}\cos^{2}\theta+a_{2}^{2}\sin^{2}\theta,\\
&\Xi_{1}:=1-\frac{a_{1}^{2}}{L^{2}},\\
&\Xi_{2}:=1-\frac{a_{2}^{2}}{L^{2}},
\end{split}
\end{equation}
where $0 \leq\phi_1, \phi_2\leq2\pi, 0\leq \theta \leq \pi/2$. The rotational parameters a and b are constrained such that $a^2, b^2 \leq \ell^{-2}$. Also, $M$ is the mass parameter, $a_{1}$ and $a_{2}$ are spin parameters, and $r_{k}$ (where $k=+, -, 0$) describe the horizon radii.
We assume that they are distinct and that $r_{0}^{2} < r_{-}^{2} < r_{+}^{2}$. Based on the above explanations, $r_{0}$ is purely imaginary, while $r_{-}$ and $r_{+}$ are real and satisfy $0 < r_{-} < r_{+}$. Therefore, $r_{-}$ represents the radius of the inner horizon, and $r_{+}$ represents the radius of the outer horizon. As shown in \cite{29} we will have,
\begin{equation}\label{19}
\begin{split}
&r_{0}^{2}+r_{-}^{2}+r_{+}^{2}+a_{1}^{2}+a_{2}^{2}+L^{2}=0,\\
&r_{0}^{2}r_{-}^{2}+r_{0}^{2}r_{+}^{2}+r_{-}^{2}r_{+}^{2}-a_{1}^{2}a_{2}^{2}+\big(2M-a_{1}^{2}-a_{2}^{2}\big)L^{2}=0,\\
&r_{0}^{2}r_{-}^{2}r_{+}^{2}+a_{1}^{2}a_{2}^{2}L^{2}=0.
\end{split}
\end{equation}
The angular velocities and Hawking temperatures at each event horizon, denoted by $r_n$ where $n=+,-,0$, are obtained according to \cite{29},
\begin{equation}\label{20}
\begin{split}
\Omega_{n,1}\equiv \frac{a_1\Xi_1}{r_n^2+a_1^2}, \hspace{0.3cm} \Omega_{n,2}\equiv \frac{a_2\Xi_1}{r_n^2+a_2^2},\hspace{0.3cm} T_n=\frac{r_n(r_n^2-r_i^2)(r_n^2-r_j^2)}{2\pi L^2(r_n^2+a_1^2)(r_n^2+a_2^2)},\hspace{0.3cm} i,j\neq n .
\end{split}
\end{equation}
We solve the Klein-Gordon equation \eqref{11} in the background of the $Kerr-AdS_{5}$  black hole.
Also, considering $\Psi=\psi(r)\psi(\theta) e^{-i\omega t+im_1\phi_1+im_2\phi_2}$ (where $m_1$ and $m_2$ are integers) and using the separation of variables \cite{29}, we convert the differential equation into two radial and angular parts.
\subsection{Angular section}
In the angular section, when $|a_1|\neq |a_2|$, and by changing the variables, as shown in \cite{29} we will have,
\begin{equation}\label{21}
\begin{split}
&x=\sin^2\theta,  \hspace{0.5cm} \psi(\theta)=(x-1)^{m_2/2}x^{m_1/2}(x-x_0)^{\tau/2} S(x)  \\
&x_0= \frac{\Xi_1}{\Xi_1-\Xi_2}, \hspace{0.5cm} \tau=\frac{m_1 a_1}{L} +\frac{m_2 a_2}{L}+\omega L .
\end{split}
\end{equation}
Then, as shown in \cite{29} for the Heun equation we will have,
\begin{equation}\label{22}
\begin{split}
\frac{d^2S(x)}{dx^2}+\bigg[\frac{\gamma_a}{x}+\frac{\delta_a}{x-1}+\frac{\epsilon_{a}}{x-x_0}+\bigg]\frac{dS(x)}{dx}+\frac{\alpha_a\beta_a x+p_a}{x(x-1)(x-x_0)}S(x)=0,
\end{split}
\end{equation}
where
\begin{equation}\label{23}
\begin{split}
&p_a=-\frac{1}{2}\left[(m_1+m_2+m_1m_2)x_0+(m_1+1)(\tau+m_1)\right]\\
&-\frac{x_0}{4\Xi_1}\left[\omega^2L^2+m^2a_1^2-\lambda+\Xi_2(m_2^2+m_1^2-\tau^2)\right],\\
&\alpha_a=\frac{1}{2}(\tau+m_1+m_2+\sqrt{m^2L^2+4}+2), \hspace{0.3cm}    \beta_a=\frac{1}{2}(\tau+m_1+m_2-\sqrt{m^2L^2+4}+2),\\
&\epsilon_a=1+\tau, \hspace{0.4cm}   \delta_a=1+m_2,  \hspace{0.4cm}    \gamma_a=1+m_1 .
\end{split}
\end{equation}
Now, we can rewrite equation \eqref{22} as follows,
\begin{equation}\label{24}
\begin{split}
&(x^3-x^2(x_0+1)+xx_0)S''(x)\\
&+[(\gamma_a+\delta_a+\epsilon_a)x^2-(\gamma_a+x_0\gamma_a+x_0\delta_a+\epsilon_a)x+x_0\gamma_a]S'(x)\\
&+[\alpha_a\beta_a+p_a]S(x)=0 .
\end{split}
\end{equation}
Therefore, by comparing relations \eqref{22}, \eqref{5}, \eqref{4}, and \eqref{3}, we obtain values for $\alpha$, $\beta$, and $a_i$.
\begin{equation}\label{25}
\begin{split}
&a_0= 1, \hspace{0.5cm} a_1=-(1+x_0), \hspace{0.5cm} a_2=x_0, \hspace{0.5cm} a_3=0\\
&a_4=\gamma_a+\delta_a+\epsilon_a, \hspace{0.5cm} a_5=-(\gamma_a+x_0\gamma_a+x_0\delta_a+\epsilon_a), \hspace{0.5cm}  a_6=\gamma_ax_0\\
&a_7=\alpha_a \beta_a, \hspace{0.5cm}  a_8=p_a, \hspace{0.5cm} \alpha=\alpha_a, \hspace{0.5cm} \beta=\beta_a .
\end{split}
\end{equation}
Also, by using relations \eqref{25} and \eqref{7}, we obtain the operators $P^+$, $P^-$, $P^0$, and $F(x\frac{d}{dx})$
\begin{equation}\label{26}
\begin{split}
&P^0=\alpha_a x \frac{d}{dx}+\beta_a\\
&P^+= x^3 \frac{d^2}{dx^2}+(\gamma_a+\delta_a+\epsilon_a)x^2\frac{d}{dx}+\alpha_a \beta_a x \\
&P^-=x_0x \frac{d^2}{dx^2}+\gamma_ax_0\frac{d}{dx}  \\
&F(x\frac{d}{dx})=-(1+x_0)x^2\frac{d^2}{dx^2}-(\gamma_a+x_0\gamma_a+x_0\delta_a+\epsilon_a)x\frac{d}{dx}+p_a .
\end{split}
\end{equation}
So, we will have according to equation \eqref{77},
\begin{equation}\label{2626}
\begin{split}
&\alpha =\frac{4 x_0}{x_0+1},   \hspace{2cm}   \beta =-\frac{x_0 \alpha _a \beta _a \gamma _a}{p_a}\\
& -\alpha  \gamma _a-\alpha  \epsilon _a-x_0 \left((\alpha -6) \gamma _a+(\alpha -3) \delta _a-3 \epsilon _a+2 \alpha +\beta -6\right)-2 \alpha -\beta=0\\
& \alpha  p_a-(\alpha +\beta ) \left(\gamma _a+\epsilon _a\right)\\
&+x_0 \left(-\gamma _a \left(-2 \delta _a-2 \epsilon _a+\alpha +\beta -2\right)+2 \alpha _a \beta _a-\alpha  \delta _a-\beta  \delta _a+2 \gamma _a^2+2 \delta _a+2 \epsilon _a\right) =0\\ .
\end{split}
\end{equation}
Since the above operators obey relation \eqref{6}, we can say that there is generalized $sl(2)$ symmetry for the angular part of the scalar field near the $ker-AdS_5$ black hole.
\subsection{Radial section}
In the radial section, the following transformation is applied \cite{29},
\begin{equation}\label{27}
\begin{split}
&z=\frac{r^2-r_+^2}{r^2-r_0^2},  \hspace{0.5cm} \psi(r)=(z-1)^{\sigma/2}(z-z_0)^{\theta_-/2}z^{\theta_+/2} R(z)  \\
&z_0= \frac{r_-^2-r_+^2}{r_-^2-r_0^2}, \hspace{0.5cm} \theta_n=\frac{i}{2\pi T_n}(\omega-m_1\Omega_{k,1}-m_2\Omega_{k,2}), \hspace{0.5cm} \sigma=2+\sqrt{4+m^2L^2} ,
\end{split}
\end{equation}
using the above transformations, the radial part of equation will become into the form of Heun equation \cite{29},
\begin{equation}\label{28}
\begin{split}
\frac{d^2R(z)}{dz^2}+\bigg[\frac{\gamma_r}{z}+\frac{\delta_r}{z-1}+\frac{\epsilon_{r}}{z-z_0}+\bigg]\frac{dR(z)}{dz}+\frac{\alpha_r\beta_r x+p_r}{z(z-1)(z-z_0)}R(z)=0,
\end{split}
\end{equation}
where
\begin{equation}\label{29}
\begin{split}
&p_r=\frac{1}{4}\left[\frac{(\lambda+m^2r_+^2-\omega^2L^2)L^2}{r_-^2-r_0^2}+2z_0(\sigma+\sigma \theta_+-\theta_+)+(\theta_-+\theta_++2)(\theta_-+\theta_+)-\theta_0^2\right]\\
&\alpha_r=\frac{1}{2}(\sigma+\theta_0+\theta_-+\theta_+), \hspace{0.3cm}    \beta_r=\frac{1}{2}(\sigma-\theta_0+\theta_-+\theta_+),\\
&\epsilon_r=\theta_-+1, \hspace{0.4cm}   \delta_r=\sigma-1,  \hspace{0.4cm}    \gamma_r=\theta_++1 .
\end{split}
\end{equation}
Note that equation \eqref{2} holds. We rewrite equation \eqref{28} as equation \eqref{2}
\begin{equation}\label{30}
\begin{split}
&(z^3-z^2(z_0+1)+zz_0)R''(z)\\
&+[(\gamma_r+\delta_r+\epsilon_r)z^2-(\gamma_r+z_0\gamma_r+z_0\delta_r+\epsilon_r)z+z_0\gamma_r]R'(z)\\
&+[\alpha_r\beta_r+p_r]R(r)=0 .
\end{split}
\end{equation}
Now, by comparing relations \eqref{30}, \eqref{5}, \eqref{4} and \eqref{3} we get
\begin{equation}\label{31}
\begin{split}
&a_0= 1, \hspace{0.5cm} a_1=-(1+z_0), \hspace{0.5cm} a_2=z_0, \hspace{0.5cm} a_3=0\\
&a_4=\gamma_r+\delta_r+\epsilon_r, \hspace{0.5cm} a_5=-(\gamma_r+z_0\gamma_r+z_0\delta_r+\epsilon_r), \hspace{0.5cm}  a_6=\gamma_rz_0\\
&a_7=\alpha_r \beta_r, \hspace{0.5cm}  a_8=p_r, \hspace{0.5cm} \alpha=\alpha_r, \hspace{0.5cm} \beta=\beta_r .
\end{split}
\end{equation}
Therefore, using relations \eqref{31} and \eqref{7}, we obtain the following operators,
\begin{equation}\label{32}
\begin{split}
&P^0=\alpha_r z \frac{d}{dz}+\beta_r\\
&P^+= z^3 \frac{d^2}{dz^2}+(\gamma_r+\delta_r+\epsilon_r)z^2\frac{d}{dz}+\alpha_r \beta_r z \\
&P^-=z_0z \frac{d^2}{dz^2}+\gamma_rz_0\frac{d}{dz}  \\
&F(z\frac{d}{dz})=-(1+z_0)z^2\frac{d^2}{dz^2}-(\gamma_r+z_0\gamma_r+z_0\delta_r+\epsilon_r)z\frac{d}{dz}+p_r .
\end{split}
\end{equation}
Here, we can obtain,
\begin{equation}\label{3232}
\begin{split}
&\alpha =\frac{4 z_0}{z_0+1},   \hspace{2cm}   \beta =-\frac{z_0 \alpha _r \beta _r \gamma _r}{p_r}\\
& -2 \alpha -\beta -\gamma _r\alpha  -\alpha  \epsilon _r-z_0 \left(2 \alpha +\beta +(\alpha -6) \gamma _r+(\alpha -3) \delta _r-3 \epsilon _r-6\right)=0\\
& \alpha  p_r-(\alpha +\beta ) \left(\gamma _r+\epsilon _r\right)\\
&+ z_0 \left(-\gamma _r \left(\alpha +\beta -2 \delta _r-2 \epsilon _r-2\right)+2 \alpha _r \beta _r-\alpha  \delta _r-\beta  \delta _r+2 \gamma _r^2+2 \delta _r+2 \epsilon _r\right)=0 .\\
\end{split}
\end{equation}

Note that the operators mentioned above are valid with respect to relation \eqref{6}. Therefore, we can conclude that the radial part of the scalar field near the Kerr black hole follows the generalized $sl(2)$ algebra. Our research reveals that the scalar field near the $RN-AdS_5$ black hole only adheres to the generalized $sl(2)$ algebra in the radial part, whereas near the $Kerr-AdS_5$ black hole, it follows the generalized $sl(2)$ algebra in both the radial and angular parts.
\section{Concluding remarks}
In the investigation of hidden symmetries in black holes, conventional methods are either based on a purely algebraic and geometric approach or on the method of finding symmetries based on the Killing vector, which has a more dominant physical approach \cite{1,2,3,4,5}. However, each of these methods has its complexities. As the metric function and geometric structure of the system become more complex, these methods may lose their effectiveness due to the complexity or difficulty of the problem. By studying the method of Castro and Maloney in the search for hidden symmetries of Kerr's black hole and other works based on this method, as well as articles that did not seek to find symmetry in cosmic structures, we realized that Klein Gordon's equations with a black hole background often become differential equations that can be solved with orthogonal polynomials.
We also know that studies of Schrödinger equations in quasi-exact solvable models have shown a logical relationship between orthogonal polynomials and the method of solving these models. The similarity between the Klein-Gordon and Schrödinger equations on one hand, and the properties of orthogonal polynomials on the other, led us to consider whether the differential equation resulting from the system, which has been transformed into a special and well-known form (Heun's equation) according to the orthogonal polynomials, can be a good candidate for directly finding the hidden algebra of the system. To this end, we selected two black holes in $AdS_5$, namely R-N and Kerr, and analyzed the Klein-Gordon equation with the background of these black holes.
As we have seen that the Klein-Gordon differential equations in the background of Reissner-Nordström and Kerr black holes can be transformed into the general form of Heun's equation by using a suitable change of variable, both in the radial and angular parts. As a result, we can easily achieve the algebraic structure based on the algebraic structure that governs the Heun equation and its coefficients. We selected the general form of Heun and demonstrated that the same results can be obtained more simply and efficiently based on the structure of Heun's coefficients, which is a shorter method than the alternative. Our work is accurate as it matches the previous works precisely.
Another interesting point is that in the work of Castro and Maloney, they were forced to impose restrictions on the omega function to remove powers higher than 2 and divide the space into two parts - near and far \cite{1}. Unlike any other application of this approximation for the Castro and Maloney method, it seems that no preconditions need to be applied on the way to the algebra in our method. This method, in addition to being more concise and practical, is also more general and comprehensive across the entire space.\\
Finally, it may be worthwhile to ask what the purpose and benefit of finding hidden symmetries, such as sl(2), for the studied systems are and what new perspectives they offer for further research? To answer this question briefly, we can say that:
The sl(2) symmetry implies that the wave equations with a black hole background can be related to a two-dimensional conformal field theory (CFT) via the AdS/CFT correspondence. This means that the black hole physics can be described by a dual CFT that lives on the boundary of the AdS space. The sl(2) generators act as the Virasoro operators of the CFT and the Casimir operator corresponds to the central charge. The sl(2) symmetry also enables us to find the eigenvalues and eigenfunctions of the Heun equation using the representation theory of the algebra \cite{39,40}.
However, In this article, our goal is not to explore these properties, but to demonstrate that we can use the wave equations with a black hole background to reveal the hidden symmetries in a system, without resorting to the common methods that may sometimes be complicated.\\
In addition to all the mentioned concepts,we also suggest some questions that can serve as motivations for further research.\\
Could renowned variants of  Heun's equation, such as the Trigonometric, Jacobi’s Elliptic,Confluent, or Weierstrass’s Form \cite{41}, be employed by reconfiguring the coefficients to unveil concealed algebras and symmetries inherent in the system?\\
Finally, this fundamental and historical question: if it is possible to solve the differential equations of a physical system using orthogonal polynomials, is it possible to point out the existence of a hidden algebra in the system based on the algebraic structure of these polynomials?
\newpage
\section{Appendix A: Calculations related to generator's communication}
In this section, we want to calculate the equation \eqref{7}, for a general function $\psi(x)$, to obtain the necessary conditions for establishing this relationship\cite{2401,24a,24b}.
\begin{equation}\label{a2}
\begin{split}
[P^+, P^-] \psi(x) =F P^0 \psi(x).
\end{split}
\end{equation}
First from the left side we will have
\begin{equation}\label{a3}
\begin{split}
&p^+ p^- \psi(x)=a_7  a_2 x^2 \psi''(x)+a_7a_6 x\psi'(x)  \\
&+a_4  a_2 x^3 \psi^{(3)}(x)+a_4a_2 x^2\psi''(x)+a_4a_6 x^2\psi''(x)     \\
& +a_0 a_2 x^4 \psi^{(4)}(x)+2a_0 a_2 x^3\psi^{(3)}(x)+a_0a_6x^3 \psi^{(3)}(x),     \\
\end{split}
\end{equation}
and
\begin{equation}\label{a4}
\begin{split}
&p^- p^+ \psi(x)= 2 a_2 a_7 x \psi'(x)+a_6 a_7 x \psi'(x)+2 a_2 a_4 x \psi'(x)+2 a_4 a_6 x \psi'(x)+a_6 a_7 \psi(x) \\
& +6 a_0 a_2 x^2 \psi''(x)+4 a_2 a_4 x^2 \psi''(x)+3 a_0 a_6 x^2 \psi''(x)+a_4 a_6 x^2 \psi''(x)+a_2 a_7 x^2 \psi''(x) \\
& + a_0 a_2 x^4 \psi^{(4)}(x)+6 a_0 a_2 x^3 \psi^{(3)}(x)+a_2 a_4 x^3 \psi^{(3)}(x)+a_0 a_6 x^3 \psi^{(3)}(x) . \\
\end{split}
\end{equation}
Using the equations \eqref{a3} and \eqref{a4}, we have
\begin{equation}\label{a5}
\begin{split}
&(p^+ p^- - p^- p^+) \psi(x) = (-4 a_0 a_2 x^3   \psi^{(3)}(x) -\left(6 a_0 a_2 +3 a_2 a_4 +3 a_0 a_6\right) x^2 \psi''(x)\\
& -\left(2 a_2 a_4 +2 a_6 a_4 +2 a_2 a_7 \right)x \psi'(x)  -a_6 a_7 \psi(x).)
\end{split}
\end{equation}
Now, we rewrite the relation \eqref{a5} as below
\begin{equation}\label{a6}
\begin{split}
&(p^+ p^- - p^- p^+) \psi(x) = \big(-4 a_0 a_2 x^3   \frac{d^3}{dx^3} \\
& -\left(6 a_0 a_2 +3 a_2 a_4 +3 a_0 a_6\right) x^2 \frac{d^2}{dx^2} \\
& -\left(2 a_2 a_4 +2 a_6 a_4 +2 a_2 a_7 \right)x \frac{d}{dx}  \\
&-a_6 a_7 \big)\psi(x).\\
\end{split}
\end{equation}
From  the righthand side of equation (37) we will also have
\begin{equation}\label{a7}
\begin{split}
&F p^0 \psi(x) = \big(\alpha  a_1 x^3  \frac{d^3}{dx^3} \\
& +\left(2 \alpha  a_1+\alpha  a_5+a_1 \beta \right)x^2 \frac{d^2}{dx^2} \\
& +\left(\alpha  a_5+\alpha  a_8+a_5 \beta \right)x \frac{d}{dx}  \\
&+a_8 \beta   \big)\psi(x),\\
\end{split}
\end{equation}
with the comparison of equations\eqref{a6} and \eqref{a7} it will be concluded that
\begin{equation}\label{a8}
\begin{split}
&\alpha =-\frac{4 a_0 a_2}{a_1},   \hspace{2cm}   \beta =-\frac{a_6 a_7}{a_8}\\
& 2 \alpha  a_1+\alpha  a_5+a_1 \beta =-\left(6 a_0 a_2+3 a_4 a_2+3 a_0 a_6\right)\\
& \alpha  a_5+\alpha  a_8+a_5 \beta =-\left(2 a_2 a_4+2 a_6 a_4+2 a_2 a_7\right) . \\
\end{split}
\end{equation}
Now, by using the above relations , the necessary conditions for establishing the equation \eqref{7} will be obtained.\\\\

\section*{Acknowledgments}
The authors would like to express their appreciation to Professor Adel Rezaei-Aghdam for his valuable feedback and advice, which has improved the quality and academic accuracy of this manuscript.\\\\

\end{document}